\documentclass[12pt]{iopart}

\usepackage[pdftex]{graphicx}
\usepackage{dcolumn}
\usepackage{bm}
 \expandafter\let\csname equation*\endcsname\relax
  \expandafter\let\csname endequation*\endcsname\relax
\usepackage{amsmath}

\usepackage[T1]{fontenc}
\usepackage[latin1]{inputenc}
\usepackage[english]{babel}
\usepackage{amssymb}
\usepackage{amstext}
\usepackage{relsize}
\usepackage{subfigure}
\usepackage{color}
\usepackage{ulem}

\begin{document}

\title{Many body population trapping in ultracold dipolar gases }

\author{Omjyoti Dutta$^1$, Maciej Lewenstein$^{2,3}$, and Jakub Zakrzewski$^{1,4}$}

\address{$^1$ Instytut Fizyki imienia Mariana Smoluchowskiego,
Uniwersytet Jagiello\'nski, ulica Reymonta 4, PL-30-059 Krak\'ow, Poland}

\address{$^2$ ICFO -- Institut de Ci\`encies Fot\`oniques,
Mediterranean Technology Park, E-08860 Castelldefels (Barcelona), Spain}
\address{$^3$ ICREA -- Instituci\'{o} Catalana de Recerca i Estudis Avan\c{c}ats, E-08010 Barcelona, Spain}

\address{$^4$ Mark Kac Complex Systems Research Center,
Uniwersytet Jagiello\'nski, Krak\'ow, Poland}
\date{\today}
\begin{abstract}


A system of interacting dipoles is of paramount importance for understanding of many-body physics. The interaction between dipoles is {\it anisotropic} and {\it long-range}. While the former allows to observe rich effects due to different geometries of the system, long-range ($1/r^3$) interactions lead to strong correlations between dipoles and frustration. In effect, interacting dipoles in a lattice form a paradigmatic system  with strong correlations and  exotic properties with possible applications in quantum information technologies, and as quantum simulators of  condensed matter physics, material science,  etc.  Notably, such a system is extremely difficult to model  due to a proliferation of interaction induced multi-band excitations for sufficiently strong dipole-dipole interactions.  In this article we develop a consistent theoretical model of interacting polar molecules in a lattice by applying the concepts and ideas of ionization theory which allows us to include highly excited Bloch bands. 
Additionally, by involving concepts from quantum optics (population trapping), we show that one can induce frustration and engineer exotic states,  such as Majumdar-Ghosh state, or vector-chiral states in such a system.
\end{abstract}

\maketitle

\section{Introduction}

In recent years the ultracold gases have been used as a tool to quantum engineer various novel  states of matter with an unprecedented precision and control. In this regard, particularly challenging is the engineering of frustrated systems for ultracold gases trapped in optical lattices. Frustration can either be induced by the lattice geometry, which can lead to kinetic frustration, or by higher order exchange processes due to strong interactions \cite{mila, macbook}. Polar molecules are particularly interesting in this context, as they can interact via long-range dipolar forces, which can induce yet another kind of frustration. In particular, dipolar lattice gases have been proposed to simulate various quantum phases and exotic phenomena, such as supersolidity \cite{maciej1, guido}, quantum magnetism \cite{gorshkov1}, topological states \cite{gorshkov2, dutta1}, exotic pair-superfluidity \cite{dutta2}, etc.  Experimental progress towards creation of quantum degenerate gas of ground state polar molecules 
has been spectacular over the last years \cite{Ni, Ospel, Aikawa, Deiglmayr}, leading, for instance,  to realization of quantum spin models  using fermionic molecules \cite{Ye} or dipolar Chromium atoms \cite{Sant}. 

One of the important properties of the polar molecules is that their dipole moment can be tuned by applying an electric field. The more polarized these molecules get, the stronger becomes the dipolar interaction between them. Theoretically it is a challenge to investigate the properties of these strongly interacting molecules trapped in an optical lattice. The standard approach based on Bose-Hubbard models limited to the lowest Bloch band  \cite{maciej1, guido,gorshkov1, gorshkov2,dutta1,dutta2} becomes inapplicable due to strong interaction induced coupling between the bands. In this paper we provide a novel route to describe such strongly interacting systems. Specifically, we consider bosonic polar molecules trapped in a one dimensional optical lattice. We find that the system can be modeled with effective couplings between the localized states at lattices sites and the continuum of highly excited states. This connects our approach to the extensive studies of strong laser field induced ionization of atoms 
and molecules. In particular we find analogies to  auto-
ionization processes, in which multi-configuration interactions couple discrete 
states with continua, as in the celebrated Fano model \cite{Fano}. Usually, due to the coupling to the continuum, the electrons in atoms or molecules are transferred from the bound states to the continuum, which leads  in the long-time limit to the irreversible decay of bound state population. Strong laser field, however, enables efficient couplings between different ionization paths leading to various interference phenomena. For strong field auto-ionization it may lead to the so called {\it confluence of coherence} \cite{Rza, Lamb}, which  slows down very efficiently  the ionization process. 

Similarly, if several (at least two) bound states are coupled to a common continuum, a phenomenon of coherent {\it population trapping} may occur -- the ionization is incomplete and a significant part of the system population is trapped in the bound subspace \cite{night}. The resulting stable bound configuration is a superposition of original bound states with properties depending on the details of the coupling to the continua. 
The coherent population trapping phenomenon appears also for multi-level discrete systems when coherent driving may create  non-absorbing states (often called ``dark states'') -- for a review of coherent population trapping see \cite{arimondo}. Most importantly, in our system of polar molecules, we find that similar phenomenon can give rise to frustration in lattice systems, as the  population trapping can involve particles trapped in {\it different} sites of the optical lattice. Specifically, we find that for a half-filling, the many-body  population trapped state is a dimer state known in the condensed-matter physics as Majumdar-Ghosh state \cite{mg}. Majumdar-Ghosh state is a paradigmatic example in the study of frustrated models, since it retains basic properties of spin-liquid phases, such as fractional excitations \cite{diep}. For lower filling we find that the effective model can be written as a $J_1-J_2$ Hamiltonian with nearest and next-nearest neighbour tunneling, along with the long range dipolar 
interactions. Similar models,  restricted only to nearest and next-nearest neighbor tunneling, have been investigated for long in connection with various magnetic materials \cite{diep}. But, in solid-state materials \cite{faze} as well as in optical lattices \cite{macbook}, such next-nearest neighbour tunneling can only come from higher order exchange processes, which makes it considerably weaker than the nearest neighbour tunneling.  The corresponding temperature is thus very low. Amazingly,  the temperature scale associated with  population trapped frustration remains comparable to the characteristic temperature scale of the system.  An alternative way to achieve long range "tunneling" in spin models is offered in  ultracold ions setting \cite{Porras-Cirac,Hauke,Maik}, but such systems are not easily scalable to macro- or even meso-scopic sizes.

\section{The model}
We consider bosonic polar molecules trapped in an optical potential inducing a one-dimensional lattice geometry, 
\begin{equation} \label{1dpot}
V_{\rm latt} = V_0\sin^2\frac{\pi x}{a}+ \frac{1}{2}m\Omega^2(y^2+z^2),
\end{equation}
where $V_0$ denotes the lattice depth and $a$ is the lattice constant. $\Omega$ denotes the harmonic (strong) trapping frequency along the $y$ and $z$ direction. The molecules are polarized by an electric field along the $z$ axis. To describe this system we make two assumptions: i) along the trapping directions, only the lowest harmonic oscillator eigenstate is occupied, and ii) at time $t<0$,  repulsively bound pairs of molecules in the limit of weak dipolar strength are prepared by tuning the lattice depth \cite{Winkler}, or by applying a weak electric field. The molecules of the pair repel each other and cannot separate due to the energy conservation - a separation would imply populating single particle states in the band gap.  Then at $t=0$, we switch on a strong polarizing electric field to induce a strong dipolar interaction between the molecules. The strength of the dipolar interaction in dimensionless units is denoted by $D=m_b \mu_{\rm ind}^2/2\epsilon_0\hbar^2a$, where $\mu_{\rm ind}$ is the 
effective dipole moment controlled by the external electric field, $\epsilon_0$ is the vacuum permittivity and $m_b$ is the mass of the molecules. The dipolar interaction is given by 
$$
V_{\rm dd}(\mathbf{r})=D\left[1-3z^2/r^2\right]/r^3. 
$$
To describe the strongly interacting regime for our system, one needs to go beyond the simple single band tight-binding approximations \cite{Jaksch}. The single-particle motion in a periodic potential results in energy bands, known as Bloch bands which can be expressed in terms of quasi-momentum $q$. For each such a band, one constructs localized basis states or orbitals [the so called Wannier functions(WF)] from the Bloch states \cite{Kohn}. By taking into account only the lowest energy Wannier states, one arrives at a Hamiltonian containing density-density interactions terms (both on-site and long-range) and nearest neighbour tunneling processes along the $x$ direction. In the presence of strong interactions such an approximation breaks down due to two primary reasons: i) the interaction mixes different bands or orbitals and different sites (specially for the higher orbitals), and ii) for higher orbitals, one has to take into account long-range tunneling matrix elements. 

These problems have been partially addressed taking into account higher excited bands in the tight-binding approximation and considering only the onsite interactions \cite{larson2009,johnson2009,will2010,dutta2011, jurgensen2012, luhmann2012, lacki2013}. For strong interactions though, serious complications appear due to the  lack of convergence of results as a function of number of bands taken into consideration. Subsequently, standard approaches become questionable and impractical. The effective description that interaction effectively increase or decrease the width of the Wannier functions \cite{dutta2011} does not hold. For such strong interactions an entirely new approach is needed. The method initiated in this work paves the way for efficient description of such systems.
   
\begin{figure}
\begin{center}
\hspace{1cm}
\includegraphics[width=130mm]{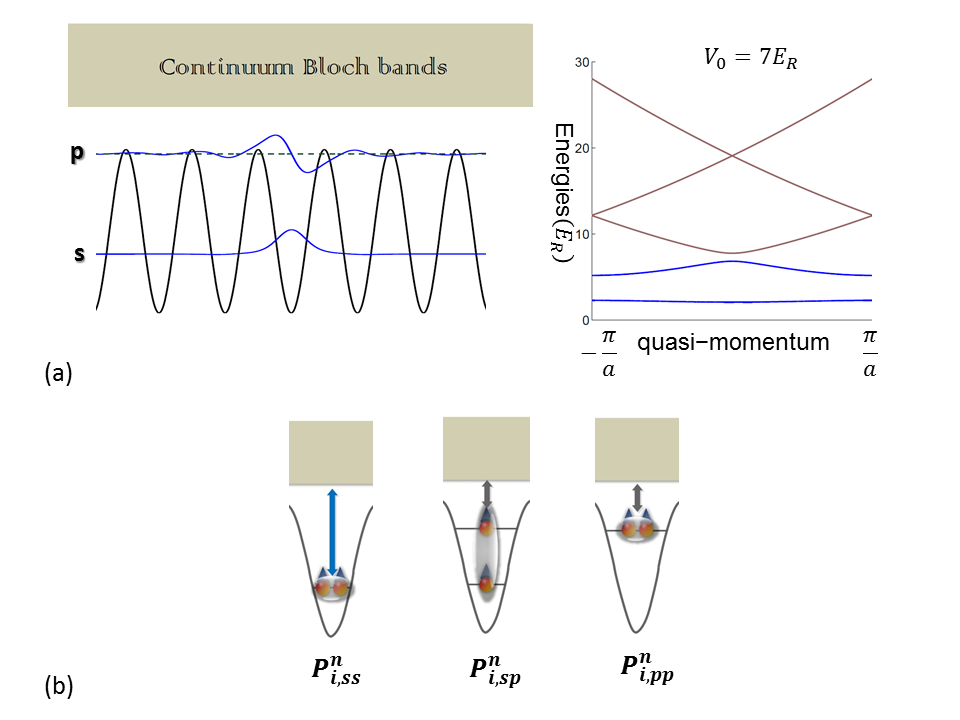}
\caption{\label{fig0} (a) The left hand side panel shows the shape and energy of the first two Wannier states (s- and p- orbitals) for lattice depth $V=7E_R$. The remaining Bloch states are represented by a continuous band of states. The plot on the right hand side shows the excitation spectrum of the Bloch bands as a function of quasi-momentum for the same lattice depth. This shows that only the first two band are separated by energy gaps whereas the higher bands form a continuum. (b) The cartoon depicts the coupling between the discrete Wanner states with two particle in the $ss$-, $sp$- and $pp$-occupied orbitals in a single site. }
\end{center}
\end{figure}

The essential observation, forming the core of our approach, is that for typical optical lattice depths, only few lowest bands are separated from each other energetically with forbidden gaps in between. The higher bands, in reality, form a continuum of energies. The simplest situation occurs
for relatively small lattice depths say of few recoil energies, as shown in
Fig.~\ref{fig0}(a) for $V_0=7E_R$. Here two lowest, $s$ and $p$, bands are separated from the continuum formed by other bands. The Wannier states (called often orbitals) of the first two bands are relatively well localized, and the mean energy calculated for them is lower than the optical lattice depth $V_0$. In this situation it is natural to express the motion of the particles in a mixed basis, where only the low-energy motion is expressed in terms of localized Wannier orbitals. Instead of using Wannier basis for the other bands too, as in the standard approaches, the remaining higher energy states will be treated by continuous Bloch functions. For much deeper lattices a natural generalization of our approach will be to take more than two discrete bands into account; in this work we limit ourselves to the simplest situation. Therefore, we may write down the field operator in the chosen mixed Wannier-Bloch basis as:
\begin{eqnarray}\label{mixedbas}
\Phi(\mathbf{r})&=&\sum_{i} \left[\hat{s}_{i}\omega^{s}_i(x)+\hat{p}_{i}\omega^{p}_i(x)\right]\phi_0(z)\phi_0(y) \nonumber\\ 
&+& \sum_{q,n>n_0} \mathcal{U}_n(q)\hat{a}_{n q}\phi_0(z)\phi_0(y)
\end{eqnarray}
where $\omega^{\mathcal{\alpha}}_i(x)$ is the localized Wannier function at site $i$ corresponding to $\alpha=s$ or  $\alpha=p$ orbital while $\phi_0$ is the lowest harmonic oscillator eigenfunction for trapping directions. $\hat{s}^\dagger_{i}, \hat{s}_{i},\hat{p}^\dagger_{i}, \hat{p}_{i}$ are the creation and annihilation operators for the bosons in the $s$- and $p$-orbitals. $\mathcal{U}_n(q)$ denotes the Bloch functions for band $n$ with quasi-momentum $q$ ($n > n_0=2$ the latter counts the number of bands treated using Wannier basis). Consequently,  $\hat{a}^\dagger_{n q}, \hat{a}_{n q}$ denote the  boson creation and annihilation operators in the bands considered in the Bloch basis with a quasi-momentum $q$. To define dimensionless quantities, we first rescale the distance $\pi x/a\rightarrow x$, that defines the scale for the energy $E_R=\pi^2\hbar^2/2m_ba^2$. So in the limit of $E_{n_0} \gg V_0$,the simplified Hamiltonian in the Wannier-Bloch basis is given by,
\begin{equation}\label{mixedham}
H = \sum_{i} E_p\hat{p}^{\dagger}_{i}\hat{p}_{i}
+H_{\rm Int}+H_{\rm Bloch} + H_{\rm WB},
\end{equation}
with
\begin{eqnarray}\label{hamint}
H_{\rm Int}&=&\sum_{i, \sigma=s,p}\frac{U_{\sigma\sigma}}{2}\hat{n}_{\sigma i}(\hat{n}_{\sigma i}-1)
+ U_{\rm ps}\sum_i\hat{n}_{si}\hat{n}_{pi}\nonumber\\ + \frac{T_{\rm ps}}{2} &\sum_i&
\left[ \hat{p}^{\dagger}_{i}\hat{p}^{\dagger}_{i}\hat{s}_{i}\hat{s}_{i} + H.c \right ] 
 +\frac{D}{2\pi^3} \sum_{\sigma,\sigma',i\ne j}\frac{\hat{n}_{\sigma i}\hat{n}_{\sigma' j}}{|i-j|^3} 
 \end{eqnarray}
 and
 \begin{eqnarray}\label{hamterm}
H_{\rm WB}&=& \sum_i\sum_{q_1 q_2; n} \left [\frac{1}{2}P^{n}_{i,ss}(q_1 q_2) \hat{a}^{\dagger}_{n q_1}\hat{a}^{\dagger}_{n q_2}\hat{s}_{i}\hat{s}_{i} + P^{n}_{i,sp}(q_1 q_2) \right . \nonumber\\ 
&\times & \left . \hat{a}^{\dagger}_{n q_1}\hat{a}^{\dagger}_{n q_2}\hat{p}_{i}\hat{s}_{i} +
\frac{1}{2} P^{n}_{i,pp}(q_1 q_2) \hat{a}^{\dagger}_{n q_1}\hat{a}^{\dagger}_{n q_2}\hat{p}_{i}\hat{p}_{i} + H.c \right ], \nonumber\\
&&
\end{eqnarray}
where $E_p$ gives the single particle energy of the $p$-orbital ($E_s=0$ is assumed). Note that, from the very begining we omit single particle tunneling terms between sites despite the lattice depth being low. That assumption is due to the fact that we shall consider a specific preparation of the system (see below) in form of pairs. The tunneling of pairs can be possible due to second order processes only. The single particle tunnelings, on the other hand, are reduced for dipoles by interaction mediated density-dependent (bond-charge) tunneling terms as discussed in \cite{dutta2}. The full Hamiltonian  is given in the Appendix A while
the estimates of the effects due to single-particle and correlated tunneling terms are discussed in Appendix B.

The interaction between particles in localized orbitals $H_{\rm Int}$ contains (with $\sigma, \sigma'= s,p$ denoting the orbitals) the onsite intra-orbital interactions $U_{\rm \sigma\sigma'}$, the possible transitions of a pair between orbitals with the strength $T_{\rm ps}$ and the long range dipolar interaction (again, additional terms in the Hamiltonian have negligible effect as discussed in the Appendix B). $H_{\rm Bloch}$ in Eq.(\ref{mixedham}) contains the kinetic energy of the molecules in the continuous band, $\sum_{q,n>n_0} E_n(q)\hat{a}^{\dagger}_{n q}\hat{a}_{n q}$ as well as interaction between particles in the continuum (see Methods section). The Wannier-Bloch Hamiltonian part $H_{\rm WB}$ describes the coupling between Wannier-described sites with two particles and the Bloch continuum. $P^{n}_{i,ss}(q_1 q_2), P^{n}_{i,sp}(q_1 q_2)$ and $P^{n}_{i,pp}(q_1 q_2)$ are the corresponding  coupling constants of two particles at site $i$ and the continuum for 
the $ss$-, $sp$- and $pp$-orbitals respectively. A cartoon of these various transition processes is shown in Fig.\ref{fig0}(b).

We would like to stress that in Eq. \eqref{hamint} we have taken into account the contribution from the dipolar interactions only. There are additional Van-der Waals terms arising from 
the mixing of rovibrational levels of molecules. Such contributions can potentially lead to a formation of long-lived molecular complexes as described in Ref.\cite{Bohn} for RbCs molecules resulting in additional loss processes which will limit the density of molecules in a lattice. Though for molecules with low density of bound molecule-molecule states such loss rate can be considerably lower.

To simplify the notation we denote the basis states for zero or two particles on a site as
\begin{equation}
\label{states}
\left |00\right \rangle\rightarrow\left |0\right\rangle, \left|20\right\rangle\rightarrow\left |1\right\rangle, \left|11\right\rangle\rightarrow\left |2\right\rangle, \left |02\right\rangle\rightarrow\left|3\right\rangle,
\end{equation}
where the state $\left|n_1n_2\right>$ denotes $n_1$ particles in the $s$-orbital and $n_2$ particles at $p$-orbital. We refer to these states as Wannier states in the following sections. 

Before considering the physics generated by the postulated Hamiltonian let us mention also that we treat the molecules rather brutally, considering them as simple dipoles. In particular we neglect the rotational structure of molecular energy levels and the induced rotational level mixing (with the effective van der Waals potential)
\cite{svet}. In unfavorable situations that may lead to creation of deeply bound molecular pairs \cite{Bohn} whose large kinetic energies allows them to leave the optical lattice potential resulting in a strong loss. These effects are discussed in more detail in Appendix B, we believe that in the parameters regime discussed below these effects can be neglected.

\section{Interesting configurations}

A large variety of different situations may be considered for the model studied. Let us imagine the situation when the system is prepared (for typical weak interactions) in an insulating state, for example the Mott state. We assume that at $t=0$ we suddenly switch on the electric field which strongly polarizes the molecules inducing large dipoles along the static field direction (assumed perpendicular to the lattice). The interaction between dipoles becomes strong making the analysis of the system difficult. Whether strong interactions will destabilize the system if the interaction energy becomes comparable to binding in the lattice? May be some metastable states still survive leading to interesting effects? These are the basic questions we want to address. 

\subsection{A single pair of molecules in neighboring sites}
First we consider the simple non-trivial situation capturing the essential physics: two neighbouring sites $i$ and $j$ share a single pair localized in either of the sites. Due to the action of $H_{WB}$, states in the neighbouring sites will be coupled via transitions to the common continuum. The state of a pair distributed among sites $i$ and $j$ may be  written as,
\begin{eqnarray}
\left|\Phi\right\rangle&=& \sum^3_{l=1}C_l \left |l\right \rangle_i\left |0\right \rangle_j\left| \mathbf{0}\right \rangle +
\sum^6_{l=4}C_{l} \left |0\right \rangle_i\left |l-3\right \rangle_j\left| \mathbf{0}\right\rangle \nonumber\\ &+& \sum_{n_1 n_2; q_1q_2} \alpha^{n_1 n_2}_{q_1 q_2} \left |0\right\rangle_i\left |0\right\rangle_j \left |\mathbf{1}_{q_1}\mathbf{1}_{q_2} \right\rangle,
\end{eqnarray}
where  $\left |l\right \rangle_i$ denotes the state of the system at site $i$ [following the notation of (\ref{states})]; $\left| \mathbf{0}\right \rangle$ denotes the vacuum for the continuum and $ \left |\mathbf{1}_{q_1}\mathbf{1}_{q_2} \right\rangle$ denotes the state with both particles in the continuum corresponding to the quantum numbers $n_1q_1$ and $n_2q_2$. The time-dependent Schr\"{o}dinger equation for $\left|\Phi\right\rangle$ leads to a set of coupled equations for
probability amplitudes $C_l$, grouped in a 6-component vector $\mathbf{C}$, corresponding to discrete states, as well as for
continuum amplitudes $\alpha^{n_1 n_2}_{q_1 q_2}$.  

\begin{eqnarray}\label{onepaireq}
i\mathbf{\dot{C}}&=&\mathbf{U}_1\mathbf{C}+\sum_{n,q_1q_2}\left[\mathbf{P}^{n}_{ij, q_1 q_2}\right]\alpha^{nn}_{q_1 q_2} \nonumber\\
i{\dot{\alpha}}^{n n}_{q_1 q_2}&=&\left[ E_n(q_1)+E_n(q_2) \right ] \alpha^{n n}_{q_1 q_2} + \left[\mathbf{P}^{n}_{ij, q_1 q_2}\right]^{\dagger}\mathbf{C} \nonumber\\
&-& \frac{\pi D \Omega_{\rm eff}}{12} \sum_{q_3,q_4}{\alpha}^{n n}_{q_3 q_4}-
\frac{\pi D \Omega_{\rm eff}}{6} \sum_{n\neq n', q_3,q_4}{\alpha}^{n' n'}_{q_3 q_4}, \nonumber\\
&&
\end{eqnarray}
where $|i-j|=1$ and $\mathbf{U}_1$ is the interaction matrix between the discrete states originating from the Hamiltonian \eqref{hamint}:
\begin{equation}
\mathbf{U}_{1} = \left( \begin{array}{cc}
\mathbf{U} & \mathbf{0}  \\
\mathbf{0} & \mathbf{U} \end{array} \right)
\end{equation}
with 
\begin{equation}\mathbf{U} = \left( \begin{array}{ccc}
U_{\rm ss} & 0 & T_{\rm ps} \\
0 & E_1+U_{\rm ps} & 0 \\
T_{\rm ps} & 0 & 2E_1+U_{\rm pp} \end{array} \right).
\end{equation} 
Due to a lack of the direct coupling between the Wannier states at different sites, $\mathbf{U}_1$ is block diagonal.
The Bloch-Wannier Hamiltonian in Eq.\eqref{hamterm} will give rise to the discrete-continuum coupling array $\mathbf{P}^{n}_{ij,q_1 q_2}=[\mathbf{P}^{n}_{i,q_1 q_2}, \mathbf{P}^{n}_{j,q_1 q_2}]^{T}$. 

To find the time evolution of the pair in the continuum, we make the ansatz that ${\alpha}^{n n}_{q_1 q_2} \approx \alpha^n$. This is justified as the attractive interaction is momentum independent and much larger than the bandwidth of the each Bloch band $n$, so that the population amplitudes have weak momentum independence. Moreover, the last term in Eq.\eqref{onepaireq} denotes coupling of population amplitude of a Bloch band $n$ to that of another Bloch band $n'$. The corresponding coupling strength $\sim D$ is of the same order of magnitude as the energy difference of the nearest Bloch bands which will be strongly coupled. Accordingly, we have assumed that for the last term in Eq. \eqref{onepaireq}, $n-n'=\pm 1$ and $\alpha^n\approx\alpha^{n-1}\approx \alpha^{n+1}$. Within these approximations, one can rewrite Eq. \eqref{onepaireq} as,
\begin{eqnarray}\label{oneeq}
i{\dot{\alpha}^n}&\approx &\left[ E_n(q_1)+E_n(q_2) - \pi D \Omega_{\rm eff} \right ] \alpha^n \nonumber \\&+& \left[\mathbf{P}^{n}_{ij, q_1 q_2}\right]^{\dagger}\mathbf{C},
\end{eqnarray}
where strong dipolar interaction effectively shifts the dispersion of each Bloch band. As initially the pairs were prepared in the discrete states in the limit of weak polarizing field, by performing Laplace transform of Eq.\eqref{oneeq} we get,
\begin{eqnarray} \label{contlap}
\alpha_n(s)=-i\frac{\left[\mathbf{P}^{n}_{ij, q_1 q_2}\right]^{\dagger}\mathbf{C}}{s-\left[ E_n(q_1)+E_n(q_2) - \pi D \Omega_{\rm eff} \right ] }.
\end{eqnarray}
Then one can do similar Laplace transform for the discrete state amplitudes in Eq.\eqref{onepaireq} 
and eliminate the continuum amplitudes by Eq.\eqref{contlap}. Subsequently, in the time evolution of the discrete state amplitudes, one gets expressions like,
\begin{equation}\label{lapl}
\sum^{n_{\rm cut}}_{n>n_0,q_1q_2}\left[\mathbf{P}^{n}_{ij, q_1 q_2}\right]\frac{\left[\mathbf{P}^{n}_{ij, q_1 q_2}\right]^{\dagger}\mathbf{C}}{s-\mathbf{i}\left[ E_n(q_1)+E_n(q_2) - \pi D \Omega_{\rm eff} \right ] },
\end{equation}
where we have introduced a cut off $n_{\rm cut} \sim 20$ in the band index and $\mathbf{i}=\sqrt{-1}$. Any excitations to higher bands than $n_{\rm cut}$, will be lost due to formation of strongly bound molecular pairs (The origin of this cut off -- the abundance of sticking collisions \cite{Bohn} -- is discussed in detail in Appendix B). Due to the shift of the energy of the continuum, the minimum of the continuum energy, $E_{n_0}-\pi D \Omega_{\rm eff}/2 \ll 0 $. Then by transforming the summation over energy level $n$ to integration, one integrates over the range $-\infty \rightarrow \infty$. 

The procedure described above  takes into account the continuum-continuum transitions in a mean-field way.
In effect,  we obtain the effective coupled equations for the time evolution  of discrete Wannier states amplitudes $\mathbf{\dot{C}}=\mathcal{M}\mathbf{C}$. The coupling matrix $\mathcal{M}$ is expressed as
\begin{eqnarray}\label{decayeq1}
\mathcal{M}&=&-\left[ i\mathbf{U}_1 + \frac{\pi}{2}\left[\frac{D\Omega_{\rm eff}}{3\pi}\right ]^2 \mathbf{\Gamma}\right ], \nonumber\\
\mathbf{\Gamma}&=& \sum_n\int\int dq_1dq_2\left[\mathbf{P}^{n}_{q_1 q_2}\right]\left[\mathbf{P}^{n}_{q_1 q_2}\right]^{\dagger}, 
\end{eqnarray}
where we have introduced the decay matrix $\mathbf{\Gamma}$ and the effective trapping strength $\Omega_{\rm eff}=\hbar\Omega/2E_R$. In the expression above $\mathbf{P}^{n}_{q_1 q_2}$ is a vector of couplings of $6$ Wannier discrete states [$3$ per site - compare \eqref{states}] with the continuum. The non-zero elements linking different sites of the discrete-continuum coupling array will induce an additional effective hopping terms for the pairs from site $i$ to site $j$. One immediately notices  that in the absence of interference effects, the decay rate of each channel will be proportional to $D^2$ [compare Eq.\eqref{decayeq1}]. Thus deviation from this behaviour may serve as an indicator of important interference terms affecting the dynamics.  

\begin{figure}
\begin{center}
\hspace{1.5cm}
\includegraphics[width=130mm]{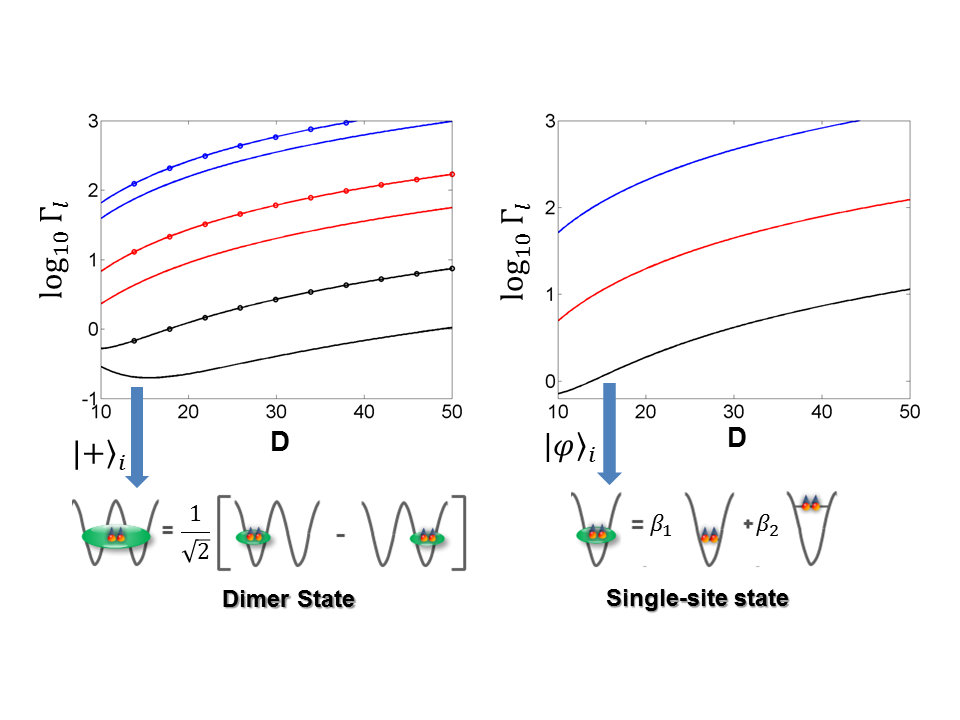}
\caption{\label{fig1} {\it The left panel}: On the top we plot the decay rates as a function of the dipolar strength $D$, when two neighbouring sites are coupled to a continuum. The pair can delocalize between the sites due to continuum-induced tunneling. Due to this coupling, each single-site channel is now decomposed into two separate channels shown by the continuous and circled lines. The state with the lowest decay rate (the black line) is described by the state,
$\left|-\right\rangle_{ij}=\left[\left|\phi\right\rangle_i-\left|\phi\right\rangle_j\right ]/\sqrt{2} $.  {\it The right panel:} Here we carried out the decay rate computations for the single-site case. On the top, we plot the decay rates as a function of the dipolar strength $D$. The decay rates are normalized to the recoil energy. The blue and the red line denotes the high decay channel with decay rates $\Gamma_l(0) \propto D^2$. The lowest decay rate channel (the black line) can be written as a superposition of two particles occupying the $s$- and $p$-orbitals as shown in the cartoon at the bottom of the figure.
}
\end{center}
\end{figure}

The full time dependent solution of the problem now reads $\mathbf{{C}}(t)=\sum^{6}_{l=1} c_{l} \exp\left[-\Gamma_l t-i\epsilon_l \right]\mathbf{u}_l$, where $\mathbf{u}_l$ is the eigenvector of the matrix $\mathcal{M}$ with  $\Gamma_l$ and $\epsilon_l$ being the decay rate and the energy of the {\it l}-th eigenstate for the neighbouring sites. 
In Fig.~\ref{fig1} (left panel) we plot the decays rates for two neighbouring sites $|i-j|=1$. 
Let us concentrate on the states with the low decay rates (the black line and the black-circled line). All the other channels (denoted by red and blue curves) have decay rates proportional to $D^2$, which points towards absence of interference effects. 
The states with low decay rates show a much different and slower scaling as a function of $D$. The corresponding eigenstates can be approximately expressed as symmetric and anti-symmetric combinations of the single-site eigenstates $ \left|\pm\right\rangle_{ij}=\left(\left|\phi\right\rangle_i\pm\left|\phi\right\rangle\right)/
\sqrt{2}$ with energies $\epsilon_{\pm}$ with $\epsilon_{+}<\epsilon_{-}$:
\begin{equation}\label{phi}
\left|\phi\right\rangle_i = \left[ \beta_1 (\hat{s}^\dagger_i)^2 + \beta_2(\hat{p}^\dagger_i)^2 \right ] \left|0\right\rangle 
\end{equation} 
expressed in terms of $s$ and $p$ orbitals.
The overlap of these approximate combinations with the exact eigenstates: $\left|\left\langle\sigma\right|\sigma'\rangle_{exact}\right|_ij\approx \mathcal{F}\delta_{\sigma\sigma'}$ is large with $\mathcal{F}\sim 0.95$, where $\sigma,\sigma'=\pm$. The deviation from the perfect overlap is due to the fact that there is an additional continuum induced off-site transition between states with opposite parity, $\left|\phi\right\rangle_i \leftrightarrow sign(i-j) \hat{s}^\dagger_j\hat{p}^\dagger_j\left|0\right\rangle $. 

The state with the lowest decay rate [the black line in Fig.\ref{fig1} (left panel)] corresponds to the state $\left|-\right\rangle_{ij}$ with highest energy. For this state we find that the ratio between the decay rate and the energy lies in the range, $\Gamma_{-}/\epsilon_{-}=0.01 \rightarrow .05$ as the dipolar strength changes from $10\rightarrow 50$. On the other hand, for the state $\left|+\right\rangle_{ij}$, for the same dipolar range, $\Gamma_{+}/\epsilon_{+}=0.05 \rightarrow 0.1$.  

It follows that on the timescale of $\sim 1/\Gamma_{+}\approx 10/E_R$, only the $\left|-\right\rangle_{ij}$ survives and will be populated. What is the origin of this surprizing stabilization? What slows down the decay in such a spectacular way? A clue lies in the fact that the analogous analysis of the fate of a pair localized in a single site only indicates a much faster decay [see Fig.\ref{fig1} -- (right panel)]. Therefore, we find a surprising situation in which a state is stabilized by delocalizing between two  neighbouring sites in the presence of continuum-induced tunneling - a coupling between sites. Such a situation is well known from single bound electron quantum optics studies -- it is the phenomenon of {\it population trapping} \cite{arimondo}. While the physics seems to be quite similar to a strong laser field induced trapping \cite{arimondo} let us stress that the ``dark state'' in our situation entangles two distinct lattice sites. We like to point out that in our scenario both the decay and 
delocalization is induced by strong coupling to the continuum. Similar analysis may be carried out for separated sites with $|i-j|>1$. It shows that in that case the effect of the continuum-assisted coupling is much smaller within the regime of dipolar strengths studied.

\subsection{Continuum-assisted creation of dimer states} 

Next we discuss the creation of dimer states due to population trapping for the half-filling of the pairs. It is known that the strong dipolar interaction induces a density-wave phase where the pairs arrange in a checkerboard pattern \cite{batrouni}. As such pairs are pinned to the sites, the checkerboard configuration will not be stable as each pair occupied site will decay rapidly to the continuum. The stable configuration can only have states containing the delocalized state $\left|-\right\rangle_{ij}$. Then, in the limit of strong interaction and for half-filling of pairs, $\left|-\right\rangle_{ij}$ will cover the whole region of lattice sites. The resulting many-body state is a checkerboard state of nearest-neighbour dimers, $\left|\Psi\right\rangle_A=\Pi_i\left| -\right\rangle_{2i,2i+1}$ or $\left|\Psi\right\rangle_B=\Pi_i\left| -\right\rangle_{2i-1,2i}$. These dimer states are the ground states of the celebrated Majumder-Ghosh (MG) model \cite{mg}. 
This paradigmatic model consists of a frustrated one-dimensional spin chain consisting of nearest and next-nearest neighbour hopping with a particular ratio. The dimer state is characterized by an absence of long-range correlations,  $\left\langle\hat{b}^{\dagger}_i\hat{b}_j\right\rangle_{\Psi}=\left\langle\hat{n}_i\hat{n}_j\right\rangle=0$ for $|i-j|>1$. This dimerized state can be thought of as the simplest form of the valance-bond solid with short-range correlations and with double the period of the original lattice. As a further support for our claim, in Appendix B, we have presented many-body calculation for small systems which shows that the state with lowest decay has almost unit overlap with the MG state. 

To prepare the MG state, initially one prepares half-filling molecular repulsively bound pairs in the regime of low dipolar interaction. Then one can switch on the strong electric field to create a strong dipolar interaction. This couples   the Wannier states to the continuum. Such a coupling usually reduces the population of molecules in the Wannier states. But in our case, due to the coherent population trapping, the initial particle density in the Wannier states will be maintained within the decay time of the population-trapped state. Any small deviation of the initial density from half-filling will manifest themselves as excitations to the final MG state. 

The doubling of periodicity in a MG state can form an experimental signature in the time of flight image due to the reduction of the Brillouin zone. The required temperature to reach this phase depends on the delocalization energy which is given by the energy difference $\delta E$ between the single-site state $\left|\phi\right\rangle_i$ and the dimer state $\left|-\right\rangle_{ij}$. For a dipolar strength of $D \sim 20$ (near the lowest decay rate in Fig.\ref{fig1}) this energy difference is of the order of $0.4E_R$. For RbCs molecules, these parameters correspond to a dipole moment of $\sim 0.7$Debye with a lattice constant $\sim 500$nm. Then the relevant temperature scale to observe this phase is $\sim 50$nK. Such a temperature is much larger than the one needed to reach the super-exchange regime for the ultracold atoms, and thus it is much easier to  access experimentally. The price to pay in our present case is the meta-stability of the dimerized state with the lifetime $\sim 10$ms. One way to 
increase the stability is by decreasing the electric field strength within the decay time, which makes all the interaction terms small. At the same time, by increasing the lattice depth one can decrease the tunneling amplitudes. This will make the dimer state frozen in time felicitating the characterization of it. 
 
  \subsection{Many-body effects due to long-range dipolar interaction} 

Let us extend our calculation of a single pair distributed in two sites to a larger system size.
We have performed an exact diagonalization for half-filled pairs distributed over $8$ sites.
Following the same procedure, we have found an effective equation of motion for the many-body discrete state probability amplitudes denoted by $\mathbf{C}_{\rm mb}$ with modified continuum induced transition matrix $\mathbf{P}_{\rm mb}$ where we have taken into account continuum induced long-range coupling. The resulting equation of motion has the form,
$\mathbf{\dot{C}}_{\rm mb}=\mathcal{M}_{\rm mb}\mathbf{C}_{\rm mb}$, and the many body coupling matrix $\mathcal{M}_{\rm mb}$  is given by 
\begin{equation}
\mathcal{M}_{\rm mb}=-\left[ i\mathbf{U}_{\rm mb} +  \frac{\pi}{2}\left[\frac{D\Omega_{\rm eff}}{3\pi}\right ]^2 \mathbf{P}_{\rm mb} \right ],
\end{equation}
where the discrete states interaction matrix $\mathbf{U}_{\rm mb}$ now also includes the long-range dipolar interaction. We then find the eigenvalues and eigenstates of the matrix $\mathcal{M}_{\rm mb}$. The real part of the eigenvalues describe the decay rate of the respective eigenstates. We then concentrate on the state with the lowest decay rate which shows similar decrease in decay strength as 
\begin{figure}
\begin{center}
\includegraphics[width=130mm]{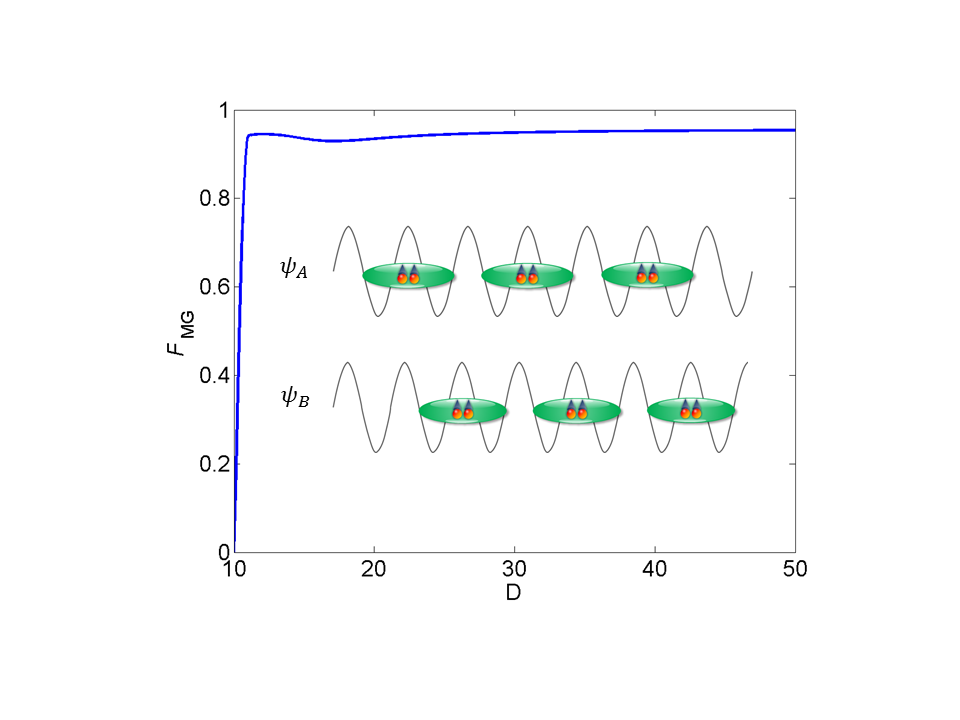}
\caption{\label{figa2} The overlap function $\mathcal{F}_{MG}$ as a function of the dipolar strength $D$.   For sufficiently large $D$, there is a large overlap with the antisymmetric MG state as defined in the text. $\Psi_{A}$ and $\Psi_B$ denotes the two configuration of the MG state.}
\end{center}
\end{figure}
the PT state discussed in the manuscript. Next, we find the overlap of this state with the Majumdar-Ghosh (MG) states ($\left|\Psi\right\rangle_A, \left|\Psi\right\rangle_B$) defined in the manuscript. We find that for larger dipolar strength $D$, there is large overlap of the lowest decay state with the antisymmetric MG state $\left|\Psi\right\rangle_{-}=\left[ \left|\Psi\right\rangle_A - \left|\Psi\right\rangle_B\right]/\sqrt{2}$. We denote this overlap by the function $\mathcal{F}_{MG}$ and plot it against the dipolar strength in Fig. \ref{figa2}. Manifestly, in spite of a strong repulsive long-range interaction between the off-site molecular pairs, the phenomenon of PT can result in a creation of the frustrated MG state. Such a result shows, furthermore,  a possibility of a creation of resonating valence-bond MG state. A detailed discussion of such a possibility is beyond the scope of the present paper.

\subsection{Constructing effective Hamiltonian for low filling}

In this section we discuss a possible way to construct an effective Hamiltonian in terms of local operators for low density of the pairs. To do this, we consider a simple system where one pair of atoms is moving in three sites coupled to the continuum. 
Following the same procedure as before we derive the full coupling matrix $\mathcal{M}(i,i+1,i+2)$ for three sites. Studying eigenstates related to the lowest decay rates we find, as before, 
that the coherent population trapping occurs due to the coupling of neighbouring sites via $\left|\pm\right\rangle_{ij}$ states. Subsequently, a tunneling Hamiltonian in terms of the states $\left|\pm\right\rangle_{ij}$ is given by,
$H_{i,i+1,i+2}= -J_{\rm eff}( \left|-\right\rangle_{i,i+1}\left\langle-\right|_{i+1,i+2}+\alpha\left|+\right\rangle_{i,i+1}\left\langle+\right|_{i+1,i+2} + h.c.) $, where 
$\alpha $ can be extracted from the eigenvalues of the effective coupling matrix $\mathcal{M}(i,i+1,i+2)$. As the states $\left|-\right\rangle_{i,i+1}, \left|\pm\right\rangle_{i+1,i+2}$ are not orthogonal, it is convenient to rewrite $H_{i,i+1,i+2}$ in terms of local orthogonal operators. To do that we define a local pair operator, $\left|\phi\right\rangle_i=b^\dagger_{i}\left|0\right\rangle$ which creates a pair at site $i$ in the lowest decay state. The pair operators satisfy bosonic commutation relations $[b_i, b^\dagger_{j}]=\delta_{ij}$. In terms of these pair operators we can rewrite the states as $\left|\pm\right\rangle_{ij}= \frac{1}{\sqrt{2}}\left[b^\dagger_{i}\pm b^\dagger_{j}\right]\left|0\right\rangle $. Subsequently, the Hamiltonian $H_{i,i+1,i+2}$ is re-expressed as,
\begin{equation}\label{Heff}
\frac{H_{i,i+1,i+2}}{J_{\rm eff}}=-\frac{1+\alpha}{2}\left[b^\dagger_{i}b_{i+1}+b^\dagger_{i+1}b_{i+2} + h.c\right] 
+(1-\alpha) b^\dagger_{i+1}b_{i+1}
+ \frac{1-\alpha}{2} \left[ b^\dagger_{i}b_{i+2}+b^\dagger_{i+2}b_{i}\right] 
\end{equation}
The values of $J_{\rm eff}$ and $\alpha$ are derived by comparing the energies of Hamiltonian \eqref{Heff} and the energies of the states with three lowest decay rates derived from the full coupling matrix $\mathcal{M}(i,i+1,i+2)$. For small values of the dipolar strength $D$ we find that $\alpha \approx 1$, thus the long-range tunneling is small and one recovers the usual picture with nearest-neighbour tunneling only. But for higher dipolar strengths $\alpha \ne 1$, due to the different decay rates of $\left|\pm\right\rangle$ states. For such values of $\alpha$ one obtains, therefore, [compare \eqref{Heff}] an effective model with next-nearest neighbour tunneling leading to frustration. We would like to point out that, in the present situation, the origin of such a frustration is entirely different from the usual origin of such terms due to the higher order processes in solid-state systems \cite{faze}.  
 
\begin{figure}
\begin{center}
\hspace{1.cm}
\includegraphics[width=130mm]{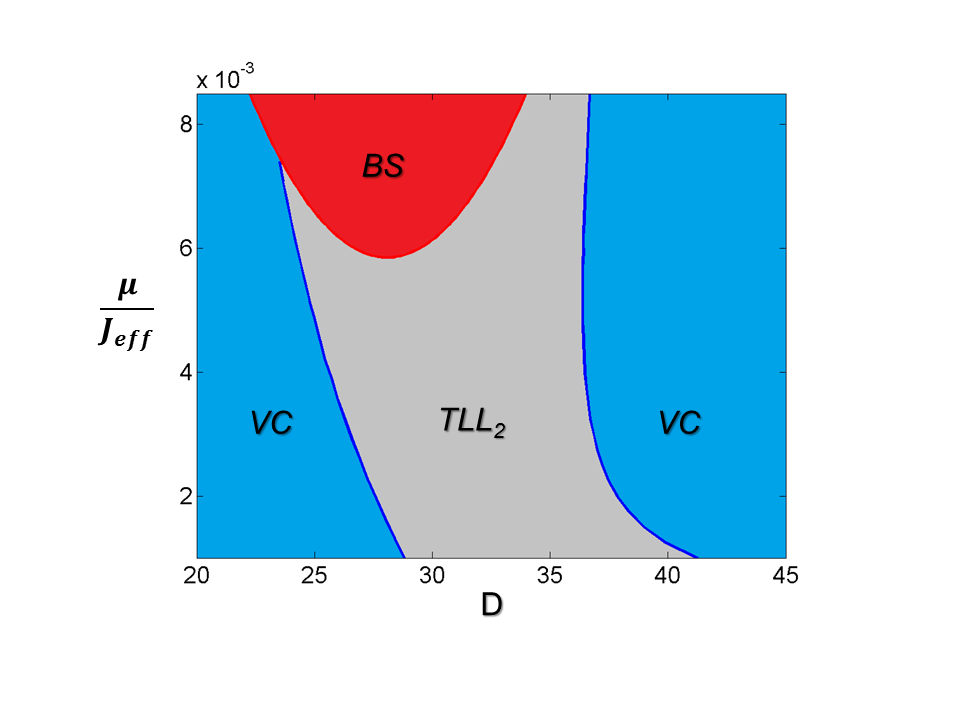}
\caption{\label{fig2} The qualitative phase diagram corresponding to the Hamiltonian \eqref{heff} as a function of dipolar strength $D$ and scaled chemical potential $\mu/J_{\rm eff}$. For low dipolar strength, the ground state is described by the vector-chiral (VC) state. With increasing dipolar strength, one finds a transition to the two-component Luttinger liquid phase (TLL$_2$) phase. A re-entrant behavior for the VC state is observed. With increasing  chemical potential, we find that the inter-component renormalized interaction $T_{12}$ becomes negative, signaling a bound state (BS) with center of mass momentum $2Q$. While such bound states are normally absent in the anti-ferromagnetic models, in our case such a situation can arise due to resonances induced by  the long-range dipolar interaction.}
\end{center}
\end{figure}

At this point, we write down the effective many-body Hamiltonian including long-range dipolar interaction and involving all sites as,
\begin{eqnarray}\label{Htot}
H_{\rm eff} &=& \sum_i H_{i,i+1,i+2} + \frac{D}{\pi^3}\sum_{ij}\frac{n_i n_j}{|i-j|^3}-\mu\sum_i n_i \nonumber\\
&=&-J_{\rm eff} (1+\alpha) \sum_{<ij>}b^{\dagger}_i b_{j}+ J_{\rm eff}\frac{1-\alpha}{2} \sum_{<<ij>>}b^\dagger_{i}b_{j}
\nonumber\\
&+& \frac{2D}{\pi^3}\sum_{i\ne j}\frac{n_in_j}{|i-j|^3} -\mu\sum_i n_i,
\end{eqnarray}
where we have introduced the chemical potential $\mu$ for the pairs and $<<ij>>$
is a shorthand for next nearest neighbour summation index. The Hamiltonian in Eq.\eqref{Htot} contains two sources of frustration: i) the effective next-nearest neighbour tunneling, and ii) long-range dipolar interaction. For our present system, the deviation of dipolar interaction from the cubic power law is negligible \cite{Carr}. The Hamiltonian, \eqref{Htot},  is a generalization of the $J_1-J_2$ model where the interaction is present to the next-nearest neighbours only. The $J_1-J_2$ model is a prototype for studying the effect of frustration and emergence of various proposed exotic phases in magnetic materials \cite{mila}. The single particle dispersion relation for this Hamiltonian is given by
$\epsilon_q=J_{\rm eff}(1+\alpha) \cos qa + J_{\rm eff} \frac{1-\alpha}{2}\cos 2qa$.  For $\frac{1-\alpha}{1+\alpha} > 1/2$ it shows two minima at wavevectors $\pm Qa=cos^{-1}\left[-\frac{1+\alpha}{2(1-\alpha)}\right]$. In our case, the two-minima limit corresponds to $D>18$. In the low-density limit, one way to treat the problem is by going to the two-component homogenous Bose gas limit \cite{vekua} with the effective Hamiltonian,
\begin{equation}\label{heff}
H_{\rm eff} =\int \left[ \frac{1}{2} T_1 (\rho^2_1 + \rho^2_2) + T_{12}\rho_1\rho_2 -\mu(\rho_1+\rho_2) \right]dx,
\end{equation}
where $\rho_{1,2}$ are the densities of the two component Bose gas centered around the the minima $\pm Q$ and $T_1, T_{12}$ are the renormalized intra-component and inter-component interaction. A detailed discussion of the Hamiltonian \eqref{heff} is presented in the methods section. For a short-range $J_1-J_2$ model, the phase diagram from such a procedure shows qualitative agreement with more involved Density-Matrix Renormalization Group simulations \cite{vekua}. When $T_1<T_{12}$, the mean-field ground state solution is given by the phase-separated state $\rho_1\ne 0, \rho_2=0$ or $\rho_1=0, \rho_2\ne 0$. Choosing one of the ground state will break the discrete symmetry which will result in true long-range order (LRO) even in one-dimension. The nature of this phase can readily be observed by writing the wavefunction in phase space, $\psi_{s}=\sqrt{\rho_s}\exp[-i\theta_s]$, with $s=1,2$. When $\rho_1=0$, we see that $\langle \hat{b}
^{\dagger}_i\rangle=\sqrt{\rho_1} \exp[-iQx+\theta]$. Such a "cone" phase is identified as a the vector-chiral (VC) phase which breaks the $\mathcal{Z}_2$ symmetry. In contrast  when $T_1>T_{12}>0$, we have a mixed state with equal density from both components. This homogeneous solution with $\rho_1=\rho_2$ is known as the two-component Tomonaga-Luttinger (TLL$_2$) liquid. There can be another possibility when the effective inter-species interaction is attractive $T_1>0, T_{12}<0$. In this situation, intra-component bound states with emerge with center of mass momentum $\sim 2Q$. Such bound states with finite momenta are usually not present in the anti-ferromagnetic model \cite{vekua}. In the present case, these bound states are a direct consequence of the long-range nature of the dipolar interaction which can induce resonances \cite{chiara}. The quasi-condensate of such bound pairs can give rise to a spin-nematic phase \cite{zhito, Sato}, or spin-density wave phase \cite{hikihara}, a detailed discussion of 
which is beyond 
the scope of 
current article. In Fig. \ref{fig2}, we have plotted  
the resulting phase diagram in the $D-\mu$ parameter space for vanishingly small $\mu$. We find that the vector-chiral phase is stable for smaller and larger values of the dipolar strength $D$. In between the homogeneous TLL$_2$ phase is the ground state. For larger values of chemical potential $\mu$, one finds that there is a bound state phase due to $T_{12}<0$.  

\section{Discussion}
Summarizing, in the present article we have demonstrated a novel approach to the problem of strongly interacting molecules in optical lattices.
We have explored a mathematical analogy between the system studied and
strong bound-continuum couplings present in the
theory of strong field ionization. We have found that the phenomenon of coherent population trapping, a well known interference effect in quantum optics, is responsible for  frustration in our system in a form of dimerization and next-nearest neighbour tunneling. One strong point of our proposal is that the required temperature scale is much higher than the one corresponding to the  usual super-exchange regime. Our results can be generalized to higher dimensions, where one can 
look for simulation of spin liquids, and valance bond crystals \cite{diep}. Our method can also be extended to other strongly interacting systems, such as  atoms in optical lattices, strongly-coupled cavity-QED systems \cite{jin}, recently proposed nano-plasmonic lattices \cite{chang}, and possible lattice geometries for the indirect excitons with strong dipolar interactions \cite{rem}. We hope that further progress can be obtained in studies of strongly interacting systems by exploring analogies with strongly coupled quantum optics problems in general, and strong field ionization theory in particular.

\section*{Acknowledgements} We thank L. Barbiero, O. J\"urgensen, D.-S. L\"uhmann, and C. Menotti for enlightening discussions. The work of O.D.\ and J.Z.\ has been supported by Polish
National Science Centre within project No. DEC-2012/04/A/ST2/00088.
M.L.  acknowledge financial support from Spanish
Government Grant TOQATA (FIS2008-01236),  EU IP SIQS, EU STREP EQuaM, and ERC Advanced
Grants QUAGATUA and OSYRIS.

\section*{Appendix A: Derivation of the Microscopic model}

The many-body Hamiltonian in terms of the field operators is given by $H=H_{0} + H_{\rm int}$,
with single particle Hamiltonian in the quasi one-dimensional optical lattice potential \eqref{1dpot}, 
\begin{equation}
H_{0}=\int d\mathbf{r} \Phi^{\dagger}(\mathbf{r})\left[-\frac{\hbar^2\nabla^2}{2m_b}+ V_{\rm latt}\right]\Phi(\mathbf{r})
\end{equation}
and the dipole-dipole interactions 
\begin{equation}H_{\rm dd}=\frac{1}{2} \int d\mathbf{r}d\mathbf{r'} \left[\Phi^{\dagger}(\mathbf{r})\Phi^{\dagger}(\mathbf{r'})V_{\rm dd}(\mathbf{r-r'})\Phi(\mathbf{r})\Phi(\mathbf{r}) \right].
\end{equation}
Representing the field operators $\Phi(\mathbf{r})$ by local site operators in the  Wannier-Bloch basis \eqref{mixedbas}
and performing appropriate integrations we find the Hamiltonian for the discrete states, discrete-continuum transitions and the continuum states. 

\subsection*{Hamiltonian for  the discrete subspace}
Here we write down the Hamiltonian originating from the single-particle kinetic energy and dipolar interaction between the discrete states,
\begin{equation}
H_{\mathrm disc}=H_{\rm T}+H_{\rm pair}+H_{\rm int}
\end{equation}
with $H_T$ describing standard and interaction induced (density-dependent) single particle tunneling
terms
\begin{eqnarray}\label{Htun}
H_{\rm T}&=& \sum_{\langle ij\rangle}\left [ -J_0\hat{s}^{\dagger}_i\hat{s}_j+J_1
\hat{p}^{\dagger}_i\hat{p}_j \right ] + \sum_{\langle ij\rangle} \left [ T_0\hat{s}^{\dagger}_i 
\left(\hat{n}_{s i}+\hat{n}_{s j}\right)\hat{s}_j+T_1
\hat{p}^{\dagger}_i\left(\hat{n}_{p i}+\hat{n}_{p j}\right)\hat{p}_j \right ] \nonumber\\
&+& \sum_{\langle ij\rangle} \left [ T_{00}\hat{s}^{\dagger}_i 
\left(\hat{n}_{p i}+\hat{n}_{p j}\right)\hat{s}_j \right.
+ \left. T_{11} \hat{p}^{\dagger}_i\left(\hat{n}_{s i}+\hat{n}_{s j}\right)\hat{p}_j \right ] 
+ T_{01}\sum_{\langle ij\rangle}{\it f}_{ij} \left[\hat{p}^{\dagger}_i\hat{n}_{s i}\hat{s}_j + h.c. \right ] \nonumber\\
 &+& T_{10}\sum_{\langle ij\rangle}{\it f}_{ij} \left[\hat{p}^{\dagger}_i\hat{n}_{p j}\hat{s}_j + h.c. \right ] 
+ {T'}_{01}\sum_{\langle ij\rangle}{\it f}_{ij} \left[\hat{p}^{\dagger}_j\hat{n}_{s i}\hat{s}_i + h.c. \right ]  \nonumber\\
&+&{T'}_{10}\sum_{\langle ij\rangle}{\it f}_{ij} \left[\hat{p}^{\dagger}_j\hat{n}_{p j}\hat{s}_i + h.c. \right ] 
\end{eqnarray}
while the correlated pair hopping part of the Hamiltonian reads
\begin{equation}
H_{\rm pair} = \sum_{\langle ij\rangle}\left[\frac{1}{2}T_{\rm p,0} \hat{s}^{\dagger}_i\hat{s}^{\dagger}_i\hat{s}_j\hat{s}_j+\frac{1}{2}T_{\rm p,1} \hat{p}^{\dagger}_i\hat{p}^{\dagger}_i\hat{p}_j\hat{p}_j + \frac{1}{2}T_{\rm p,01} \left(\hat{s}^{\dagger}_i\hat{s}^{\dagger}_i\hat{p}_j\hat{p}_j+h.c.\right)+T_{\rm p,10} \hat{s}^{\dagger}_i\hat{p}^{\dagger}_i\hat{p}_j\hat{s}_j\right]
\end{equation}
where $J_0, J_1>0$ denote the single particle nearest neighbor tunneling amplitudes in the $s,p$-orbital respectively.
Intra-orbital interaction-induced tunneling amplitudes are denoted by $T_{0}, T_{1}, T_{00}, T_{11}$. The interaction-induced inter-orbital tunneling amplitudes are given by $T_{01}, {T'}_{01}, T_{10}, {T'}_{10}$. The staggered nature of the inter-orbital tunneling is denoted by ${\it f}_{ij}=\pm 1$ when $i-j=\mp 1$. The pair tunneling Hamiltonian is denoted by $H_{\rm pair}$ and the corresponding pair-tunneling amplitudes are given by $T_{\rm p,0}, T_{\rm p,1}, T_{\rm p,01}, T_{\rm p,10}$.  
All these terms, partially canceling each other, are neglected in our simplified Hamiltonian \eqref{mixedham}. The extended analysis taking into account single particle tunneling is discussed in Appendix B, below.

Next, we rewrite the dipolar interaction between the various Wannier orbitals from Eq.(3) in the main text,
\begin{eqnarray}\label{Hint}
H_{\rm int} &=& \sum_{i, \sigma=s,p}\frac{U_{\sigma\sigma}}{2}\hat{n}_{\sigma i}(\hat{n}_{\sigma i}-1)
+ U_{\rm ps}\sum_i\hat{n}_{si}\hat{n}_{pi} \nonumber\\
&+& \frac{T_{\rm ps}}{2} \sum_i \left[ \hat{p}^{\dagger}_{i}\hat{p}^{\dagger}_{i}\hat{s}_{i}\hat{s}_{i} + H.c \right ] 
 +\frac{D}{2\pi^3} \sum_{\sigma,\sigma',i\ne j}\frac{\hat{n}_{\sigma i}\hat{n}_{\sigma' j}}{|i-j|^3} 
\end{eqnarray}
The different amplitudes in the discrete subspace Hamiltonian are obtained by appropriate integrals of the dipole-dipole interaction potential and the mode functions [compare \eqref{mixedbas}] that contain Wannier functions for orbitals along $x$ with product of ground state Gaussians in perpendicular direction. For completness we list these integrals explicitly,
assuming a shorthand notation
\begin{equation}
{\cal W}(\mathbf{r,r'})=V_{\rm dd}(\mathbf{r-r'})\phi^2_{0}(y)\phi^2_{0}(y')\phi^2_{0}(z)\phi^2_{0}(z')
\nonumber
\end{equation}
and assuming the Wannier functions to be real:
\begin{eqnarray}
U_{\rm ss}=\int &d\mathbf{r}d\mathbf{r'}& |\omega^{s}_i(x)\omega^{s}_i(x')|^2 {\cal W}(\mathbf{r,r'}), \\
U_{\rm pp}=\int &d\mathbf{r}d\mathbf{r'}&|\omega^{p}_i(x)\omega^{p}_i(x')|^2 {\cal W}(\mathbf{r,r'}),\nonumber\\
T_{0}=\int &d\mathbf{r}d\mathbf{r'}&\left[\omega^{s}_j(x)\right]^3\omega^{s}_i(x') {\cal W}(\mathbf{r,r'}),\nonumber\\
T_{1}=\int &d\mathbf{r}d\mathbf{r'}&\left[\omega^{p}_j(x)\right]^3\omega^{p}_i(x'){\cal W}(\mathbf{r,r'}), \nonumber\\
T'_{01}=\int &d\mathbf{r}d\mathbf{r'}&\omega^{p}_j(x)\omega^{s}_i(x)\left[\omega^{s}_i(x')\right]^2{\cal W}(\mathbf{r,r'}), \nonumber\\
T'_{10}=\int &d\mathbf{r}d\mathbf{r'}&\omega^{p}_j(x)\omega^{s}_i(x)\left[\omega^{p}_i(x')\right]^2 {\cal W}(\mathbf{r,r'}), \nonumber\\
T_{\rm ps}=\int &d\mathbf{r}d\mathbf{r'}&\omega^{p}_i(x)\omega^{p}_i(x')\omega^{s}_i(x)\omega^{s}_i(x') {\cal W}(\mathbf{r,r'}),\nonumber\\
T_{\rm p,0}=\int &d\mathbf{r}d\mathbf{r'}&\omega^{s}_i(x)\omega^{s}_j(x)\omega^{s}_i(x')\omega^{s}_i(x') {\cal W}(\mathbf{r,r'}),\nonumber\\
T_{\rm p,1}=\int &d\mathbf{r}d\mathbf{r'}&\omega^{p}_i(x)\omega^{p}_j(x)\omega^{p}_i(x')\omega^{p}_i(x'){\cal W}(\mathbf{r,r'}),\nonumber\\
T_{\rm p,01}=\int &d\mathbf{r}d\mathbf{r'}&\omega^{s}_i(x)\omega^{p}_j(x)\omega^{s}_i(x')\omega^{p}_j(x') {\cal W}(\mathbf{r,r'}),\nonumber\\
T_{\rm p,10}=\int &d\mathbf{r}d\mathbf{r'}&\left [ \omega^{s}_i(x)\omega^{s}_j(x)\omega^{p}_i(x')\omega^{p}_j(x') + \omega^{s}_i(x)\omega^{p}_j(x) \omega^{p}_i(x')\omega^{s}_j(x')\right]{\cal W}(\mathbf{r,r'}), \nonumber\\
U_{\rm ps}=\int &d\mathbf{r}d\mathbf{r'}&\left [ |\omega^{p}_i(x)\omega^{s}_i(x')|^2 + \omega^{p}_i(x)\omega^{s}_i(x)\omega^{p}_i(x')\omega^{s}_i(x')\right] {\cal W}(\mathbf{r,r'}),\nonumber\\
T_{00}=\int &d\mathbf{r}d\mathbf{r'}& \left[\omega^{s}_i(x)\omega^{s}_j(x)\left[\omega^{p}_i(x')\right]^2 + \omega^{s}_i(x)\omega^{p}_i(x)\omega^{p}_i(x')\omega^{s}_j(x')\right] {\cal W}(\mathbf{r,r'}), \nonumber\\
T_{11}=\int &d\mathbf{r}d\mathbf{r'}& \left[\omega^{p}_i(x)\omega^{p}_j(x)\left[\omega^{s}_i(x')\right]^2 + \omega^{p}_i(x)\omega^{s}_i(x)\omega^{s}_i(x')\omega^{p}_j(x')\right] {\cal W}(\mathbf{r,r'}), \nonumber\\
T_{01}=\int &d\mathbf{r}d\mathbf{r'}& \left[\omega^{p}_i(x)\omega^{s}_j(x)\left[\omega^{s}_i(x')\right]^2 + \omega^{p}_i(x)\omega^{s}_i(x)\omega^{s}_i(x')\omega^{s}_j(x')\right]{\cal W}(\mathbf{r,r'}), \nonumber\\
T_{10}=\int &d\mathbf{r}d\mathbf{r'}& \left[\omega^{p}_i(x)\omega^{s}_j(x)\left[\omega^{s}_j(x')\right]^2 + \omega^{p}_i(x)\omega^{p}_j(x)\omega^{p}_j(x')\omega^{s}_j(x')\right]{\cal W}(\mathbf{r,r'}) , \nonumber\\
\end{eqnarray}

\subsection*{The continuum states and their couplings to bounded subspace}
 Next  we consider relevant  properties of Bloch states. Let us denote the Bloch band $n_0$ as the start of the continuous bands with $E_{n_0}$ as the minimum of energy and $E_{n_0}\gg 1$. Then the energy of the Bloch band $n=n_0+m$ can be written as, $E_{n=n_0+m}(q)=E_{n_0}+2n_0m+m^2+2(n_0+m)|q|+q^2$ for even $m$ and one can get similar results for 
odd $m$. Moreover, we  write the Bloch wavefunctions in the $E_{n_0}\gg 1$ limit as \cite{Mathieu},
$\mathcal{U}_{n q}(x)\approx\sqrt{\frac{2}{L}}\exp[i q x]\cos\left[ \sqrt{E_n-q.^2-V/2}x \right], $
for even  $n$ and $\mathcal{U}_{n q}(x)\approx \sqrt{\frac{2}{L}}\exp[i q x]\sin\left[ \sqrt{E_n-q.^2-V/2}x \right]$ for $n$ odd. As these functions are eigenstates, they are also orthogonal, $ \int U^{*}_{n q}(x)U^{*}_{m q'}(x) dx = \delta_{n,m}\delta_{q,q'}.$
Now we write down the discrete-continuum coupling matrix elements as,
\begin{eqnarray*}
P^{n}_{i,p_1p_2}(q_1 q_2)&=& \int U^{*}_{n q_1}(x)U^{*}_{n q_2}(x')V_{\rm dd}(\mathbf{r-r'})\omega^{p_1}_i(x')\omega^{p_2}_i(x) \nonumber\\ 
&\times& |\phi_0(z)\phi_0(y)|^2|\phi_0(z')\phi_0(y')|^2 d\mathbf{r}d\mathbf{r'} , \\
&\approx & \frac{1}{4}\int dk V_{\rm dd}(k)\left [ \mathcal{W}^{\sigma_1}_i(k-q_1+\sqrt{E_{n_1}(q_1)}) \right . \nonumber\\ 
&\times & \left . \mathcal{W}^{\sigma_2}_i(-k-q_2-\sqrt{E_{n_2}(q_1)}) + \right . \nonumber\\
&& \left . \mathcal{W}^{\sigma_1}_i(k-q_1-\sqrt{E_{n_1}(q_1)}) \right . \nonumber\\
&\times& \left . \mathcal{W}^{\sigma_2}_i(-k-q_2+\sqrt{E_{n_2}(q_1)})\right ] ,
\end{eqnarray*}
where $\mathcal{W}^{\sigma}_i(k)$ is the Fourier transform of the Wannier function $\omega^{\sigma}_i(x)$.
In deriving the above form, we have used the orthogonality condition between the Bloch functions and assumed that $E_n \gg 1$. 
Additionally, in the Hamiltonian \eqref{mixedham}, we have neglected terms corresponding to  processes like $\hat{a}^{\dagger}_{n q_1}\hat{s}^{\dagger}_{i}\hat{s}_{i}\hat{s}_{i}$ where one particle is coupled to the continuum. The transition amplitudes for such processes contains
convolution sums of the form
$$
S_n\sim \int \mathcal{W}^{\sigma_1}(k+q+\sqrt{E_n})\mathcal{W}^{\sigma_2}(-k) V_{\rm dd} (k)dk.
$$
As $E_n \gg 1$, such terms are negligibly small. Thus we ignored them in comparison to the leading two-particle transition amplitudes.

\subsection*{Hamiltonian for the continuum states} 
The Hamiltonian for the continuum Bloch states reads
\begin{eqnarray}\label{hcont}
H_{\rm Bloch} &\approx & \sum_{n,q} E_n(q)\hat{a}^{\dagger}_{n q}\hat{a}_{n q} - \frac{\pi D \Omega_{\rm eff}}{12}\sum_{n,\mathbf{q}}\hat{a}^{\dagger}_{n q_1}\hat{a}^{\dagger}_{n q_2}\hat{a}_{n q_3}\hat{a}_{n q_4} \nonumber\\
&-& \frac{\pi D \Omega_{\rm eff}}{6} \sum_{n\neq n', \mathbf{q}} \hat{a}^{\dagger}_{n q_1}\hat{a}^{\dagger}_{n q_2}\hat{a}_{n' q_3}\hat{a}_{n' q_4} \nonumber\\
&-& \frac{\pi D \Omega_{\rm eff}}{6} \sum_{n \neq n', \mathbf{q}} \hat{a}^{\dagger}_{n q_1}\hat{a}^{\dagger}_{n' q_2}\hat{a}_{n q_3}\hat{a}_{n' q_4},
\end{eqnarray}
the continuous band index $n, n'>n_0$ and the momentum index $\mathbf{q}=[q_1, q_2, q_3, q_4]$.
The second term in the Hamiltonian \eqref{hcont} denotes the dipolar interaction between the molecules in the same Bloch band $n$ whereas the next term denotes the transition of pairs between two Bloch bands and the last term denotes interaction between molecules from different Bloch bands.
We only include the leading terms whose strength is of the order of $\sim D$. Furthermore, from Hamiltonian \eqref{hcont}, we notice that the interaction is strongly attractive in the higher Bloch bands and for strong interaction ($D\gg 1$), the dipolar strength can exceed the width of the first few continuous Bloch bands.

\subsection*{Many-body effects in the continuum} 

Consider  the effect of dipolar interaction when many pairs decay into the continuum. Again,
within each Bloch band, the dipolar attraction is 
larger than the respective bandwidth of the Bloch band. This suggests strong binding of the molecular pairs. To denote this we introduce a composite operator for the pairs,
$$\hat{b}^{\dagger}_n=\frac{\int\int\hat{a}^{\dagger}_{n q_1}\hat{a}^{\dagger}_{n q_2}dq_1dq_2}{\int\int dq_1dq_2}$$. 
As the molecules can scatter to any quasi-momentum state with equal strong probability, one can assume that each quasi-momentum level in the band $n$ is at most occupied by one molecule. Then, in terms sof the pairing operator,  one can find an momentum average representation Hamiltonian \eqref{hcont} in terms of the composite operators as, 
\begin{eqnarray}\label{hpair}
H_{\rm Bloch} &\approx & \sum_{n} \left [\epsilon_{n, \rm avg} -\frac{\pi D \Omega_{\rm eff}}{3} \right ] \hat{b}^{\dagger}_{n}\hat{b}_{n} - \frac{2\pi D \Omega_{\rm eff}}{3} \sum_{n\neq n'}  \hat{b}^{\dagger}_{n}\hat{b}_{n'} \nonumber\\
&-& \frac{2\pi D \Omega_{\rm eff}}{3} \sum_{n \neq n'} \hat{b}^{\dagger}_{n}\hat{b}_{n}\hat{b}^{\dagger}_{n'}\hat{b}_{n'},
\end{eqnarray}
where the average dispersion energy of a pair in Bloch band $n$ is given by $\epsilon_{n, \rm avg}=2\int E_n(q)dq $. From the Hamiltonian \eqref{hpair}, by taking a mean-field type approximation for the composite operator will again result is the effective shift in the dispersion.

\section*{Appendix B: Testing the approximations}

\subsection*{Small system analysis of a single pair}
Let us reconsider the model of a pair distributed over neighbouring sites. This time we include  the effect of pair breaking due to the single particle tunneling matrix in Hamiltonian
Eq. \eqref{Htun} and Eq.\eqref{Hint}. To do that, within the two-site model, we have reevaluated the dynamics of the pairs by taking into account states with single molecule per site. 
Our initial state consists of the situation where only one of the site contains a pair. 
\begin{figure}
\begin{center}
\includegraphics[width=80mm,natwidth=610,natheight=642]{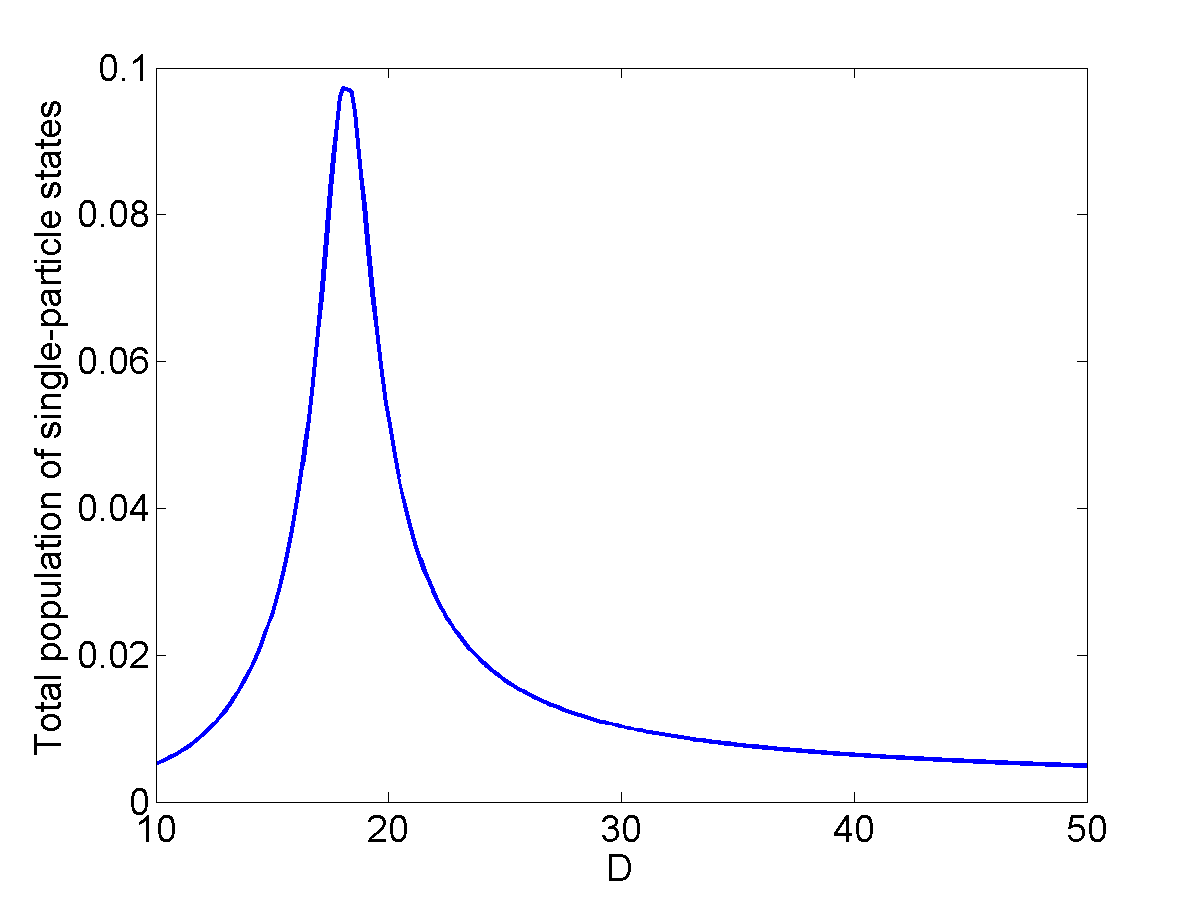}
\caption{\label{figa1} Here we plot the total population of the single particles states distributed over the two sites as a function of the dipolar strength $D$. We find that the sing-particle states have negligible population.}
\end{center}
\end{figure}
With this initial condition, we have carried out the full dynamics within the two-site case and the result is presented in Fig.\ref{figa1}. There we have plotted the total population of the single-particle states. We see that the maximum population of the single particle states are less than $< 0.1$. The main reason for such anobservation is that within the Wannier orbitals, the effective single-particle tunneling terms are much smaller (due to the aspect ratio of a site in the lattice) than the continuum induced pair tunnelings that are independent of any local aspects of the Wannier function. This justifies our assumption of neglecting  the pair-breaking effect of the single-particle tunneling Hamiltonian. Moreover, due to such a negligible population of the single-particle states, the decay rates of various channel remains unchanged with respect to the case discussed in the paper. 

\subsection*{Effect of Van der Waals (VdW) potential due to rotational level mixing} 
We discuss here  the effect of rotational level mixing due to quantum nature of the dipolar interaction, the
effect neglected in the main text. Such a mixing gives rise to an effective VdW like potential which decays with distance $r$ as $-1/r^6$ \cite{svet}. To look into its effect, we first consider a polar molecule with dipole moment $\mu$, rotational constant $B_e$ is polarized by a strong electric field $\mathbf{E}$ along the $z$ direction. In the limit of $(\mu \mathbf{E}/\hbar B_e)\ll 1$, one can write the rotational Hamiltonian in the $M=0$ sector ($M$ is the projection of angular momentum along the molecular axis) as,
$$
H_{\rm rot}=\hbar B_e \hat{J}^2-\mu \mathbf{E} \cos \theta \approx -\hbar B_e \partial^2_{\theta} + 
\mu \mathbf{E} \theta^2/2,
$$
where $\theta$ is the angle between the molecular axis and the electric field direction and $\mu$ is the permamnent dipole moment. The energy levels of the Hamiltonain $H_{\rm rot}$ is denoted by the index $m=0,1,2,.....$ with energy $E_{\rm rot, m} = (2m+1) \hbar B_e/d^2_{\theta}$ and wavefunction 
$\Phi_m(\theta)=N_m\exp(-\theta^2/2d^2_{\theta})H_m(\theta/d_{\theta})$ where $H_m(.)$ is the Hermite polynomial of order $m$, $N_m$ is the normalization constant and the width $d_{\theta}=\left[2 \hbar B_e/\mu \mathbf{E}\right ]^{1/4}$. The rotational state of the polar molecule is denoted by the lowest energy rotational wavefuntion $\Phi_0(\theta)$ which induced a dipole moment of $\mu_{\rm ind} \mu\int \cos\theta\Phi^2_0(\theta)d\theta$. This results in dipolar interaction between the ground state molecules which falls of as $1/r^3$. Additionally, dipolar interaction also induces excitations to higher energy rotational states. Within second order perturbation theory, the resulting effective interaction between the ground state 
molecules, in the units of recoil energy, is given by,
\begin{equation}\label{Van}
V_{\rm VdW}(\mathbf{r}) \approx - \frac{\ell^4_{\rm VdW}}{r^6}\left [ 1-3z^2/r^2\right ]^2, 
\end{equation}
where the distance are in the units of $a/\pi$ and the effective dimensionless VdW length $\ell_{\rm VdW}$ is given by,
$$
\ell_{\rm VdW}= \left[d^2_{\theta}\exp\left(-d^2_{\theta}/2\right)\frac{D^2_{\rm max}E_R}{4\sqrt{2}\hbar B_e}\right]^{1/4}/\pi,
$$ =
where the maximum dipolar strength is given by $D_{\rm max}=m_b\mu^2/2\epsilon_0\hbar^2 a$. For RbCs molecule, the rotational constant is given by $B_e=0.014$cm$^{-1}$ and the permanent dipole moment is given by $\mu=1.27$ Debye. Then for an electric field strength of $E=10$kV/cm, the angular width reads $d_\theta=0.7$. Correspondingly, the VdW length is given by $ \ell_{\rm VdW} \approx 0.17$ when
the lattice constant is $a=500$nm. From this we can also define a short distance cutoff scale $\ell_{\rm sr}$ where rotational mixing effect of the dipoles becomes similar magnitude to the rotational splitting \cite{svet}. In our units, this cut off is given by $\ell_{\rm sr} \approx .03$ for dipolar strength $D=20$. For length scales $r >\ell_{\rm sr}$, the perturbative form of the VdW interaction in Eq.\eqref{Van} remain valid and for $r<\ell_{\rm sr}$, the rotational level of the molecules becomes strongly mixed and the deeply bound molecular pairs appears \cite{Bohn}.

Following the discussion in the main text and the above sections, we write the VdW Hamiltonian in the discrete ($H_{\rm VdW, int}$), continuous ($H_{\rm VdW, Bloch}$) and discrete-continuous ($H_{\rm VdW, WB}$) sector. The interaction in discrete sector is weak compare to the dipolar interaction. This can be easily seen by Fourier transforming Eq. \eqref{Van}, $V_{\rm VdW}(k) \approx \ell^4_{\rm VdW} k^3 \mathcal{F}(k\ell_{\rm VdW}) $, with the function $\mathcal{F} \sim 1$. The widths of the Wannier functions in the momentum space are of the order of $k \sim 1$. Then as $\ell^4_{\rm VdW} \sim 10^{-3} \ll 1$, we can neglect the VdW interaction in the discrete states compare to the dipolar strength in Eq. \eqref{hamint}. 

Moreover, one can estimate the loss rate due to the coupling of the bound molecular complex by evaluating the overlap between the Wannier orbitals and the bound state wave function which is of the order of $\exp(-1/\ell_{\rm VdW})\rho$ where $\rho$ is the density of bound states in the units of recoil energy. Here we have assumed that the bound state decays exponentially for a large distance. From Ref.\cite{Bohn}, for RbCs molecue in the rotational ground state, the density of states is large, $\rho \sim 40$. Accordingly, the decay rate will be proportional to the overlap which is of the order of 
$0.1E_R$ which gives a timescale of $\sim 1.0$ms. For other species of molecules it is possible that the
density of bound states is lower which can result in an increased stability.

In the continuous Bloch band, the corresponding momentum scale is given by $k \sim \sqrt{E_n}$ and
the corresponding strength of the VdW interaction in the continuum band $n$ is in the order of
$-\ell^4_{\rm VdW} E^{3/2}_n $. Whereas from Eq. \eqref{hcont}, we find that the strength of the dipolar attraction in this band is of the order of $\sim D $. For dipolar strength of $D \sim 20$,
the VdW interaction gets prominent only for very high Bloch bands with $n_{\rm cut off} \gtrsim 1/\ell_{\rm sr} \approx 30$. As such bands probes distance shorter than $\ell_{\rm sr}$, this will result in strong overlap (or in other words, strong coupling) with the molecular bound states which can give rise to phenomenon of molecular sticking \cite{Bohn}. Subsequently, any population in those Bloch bands will result in loss due to formation of strongly bound molecular pairs and we denote this by loss rate $\Gamma_{\rm VdW}$ 

The situation remains similar also for the discrete to continuous transitions. There the transitions happens between states with momentum $k \sim \sqrt{E}_n$ continuous states and discrete states with momentum $k \sim 1$. As for continuous states, $E_n \gg 1$, the corresponding VdW discrete-continuous transition strength is of the order of $-\ell^4_{\rm VdW} E^{3/2}_n$. Subsequently, for intermediate momentum, the discrete-continuum transition is again dominated by the dipolar terms in Hamiltonian \eqref{hamterm}and molecular sticking due to VdW interaction involves very high energy Bloch bands (or shorter distance) with band index $n \gtrsim 30$.

Accordingly, while integrating out the high Bloch bands, one get additional terms (equivalent to the term in Eq.\eqref{lapl1}),
\begin{eqnarray}\label{lapl1}
\sum_{n>n_{\rm cut}}&&\left[\mathbf{P}^{n}_{ij, q_1 q_2}\right]\frac{\left[\mathbf{P}^{n}_{ij, q_1 q_2}\right]^{\dagger}\mathbf{C}}{s-\mathbf{i}\left[ E_n(q_1)+E_n(q_2) - \pi D \Omega_{\rm eff} \right ]+ \Gamma_{\rm VdW}}  \nonumber\\
&\approx & \tan^{-1}\left[\frac{\Gamma_{\rm VdW}}{n^2_{\rm cut}-\pi D \Omega_{\rm eff}}\right] \approx 0,
\end{eqnarray}
as $n^2_{\rm cut} \gg D $, and the decay rate $\Gamma_{\rm VdW} \sim D$ for a lattice constant of $500$nm and dipolar strength $D\sim 20$. We have calculated $\Gamma_{\rm VdW}$ from Ref.\cite{Bohn} but assuming temperature in the nano-Kelvin regime which suppresses the $d$-wave resonances.

\subsection*{Two-component Bose gas limit of Eq.\eqref{Htot} }
We rewrite our Hamiltonian \eqref{Htot} in the conventional $J_1-J_2$ form as,
\begin{eqnarray}\label{j1j2}
H_{\rm eff} &=& J_1 \sum_{<ij>}b^{\dagger}_i b_{j}+ J_2 \sum_{<<ij>>}b^\dagger_{i}b_{j}\nonumber\\
&+& V \sum_{ij}\frac{n_in_j}{|i-j|^3} -\mu\sum_i n_i,
\end{eqnarray}
where $J_1, J_2$ are the nearest and next-nearest neighbour tunneling and $V$ is the strength of the long-range interaction. In the dilute limit, such a system, with nearest and next-nearest neighbour interaction only, has been solved qualitatively by mapping the problem to a two-component Bose gas model\cite{vekua}. Here we extend this treatment to include long-range dipolar interaction. To do that we transform the Hamiltonian to the momentum space,
\begin{equation}\label{hk}
H_{\rm eff} = \sum_{q}\epsilon_q b^{\dagger}_q b_{q} + \sum_{k,k',q}V(q) b^{\dagger}_{k+q} b^{\dagger}_{k'-q} b_{k'} b_{k} -\mu\sum_i b^{\dagger}_q b_{q},
\end{equation}
where the dispersion relation is given by $\epsilon_q=2J_1\cos qa + 2J_2\cos 2qa $ and the interaction energy in momentum space is given by, $V(q)=U+2V\sum^{\infty}_{n=1} \cos nqa/n^3$, where the hard-core constraint is given by $U \rightarrow \infty $. We only consider the dilute limit, $\mu \rightarrow 0$. When $J_2 > J_1/4$, the dispersion relation has two minima at wavevectors, $Qa=\cos^{-1}\left[-J_1/4J_2 \right]$. Around these minima, we can write the dispersion relation as, $\epsilon_{Q+k}=\epsilon_{Q}+
\hbar^2 k^2/2m^{*}$, where $m^{*}$ is the effective mass. Then we expand the boson operator near the two minima, $b_k=\phi_{1,Q+k}+\phi_{2,-Q+k} + \phi_k$, where $\phi_1$ and $\phi_2$ are the two-component Bose gas centered around momentum $\pm Q$ respectively, while $\phi_k$ denotes the high momentum contribution, which is integrated out. Then one can re-express the Hamiltonian \eqref{hk} in terms of the $\phi_{1,2}$ which in position space reads,
\begin{eqnarray}
H_{\rm eff}&=&\int dx \left[ \sum_{\sigma=1,2} \left[-\phi^{\dagger}_{\sigma}\frac{\hbar^2}{2m^{*}}\nabla^2_x\right]\phi_{\sigma} \right.  \nonumber\\
&+& \left . \frac{1}{2} T_1 (\rho^2_1 + \rho^2_2) + T_{12}\rho_1\rho_2 -\mu(\rho_1+\rho_2)\right ], 
\end{eqnarray}
where $T_1$ and $T_{12}$ are renormalized interactions. To find these renormalized interactions, we first write down the full Bethe-Salpeter equation,
\begin{equation}\label{renor}
T(k,k';q)=V(q)-\int\frac{V(p-q)T(k,k';p)}{\epsilon_{k+p}+\epsilon_{k'-p}+\Omega} \frac{dp}{2\pi}.
\end{equation}
In the dilute limit we can substitute $\Omega=2\mu$. Then the respective renormalized interaction is given by, $T_1=T(Q,Q,0)$ and $T_{12}=T(Q,-Q;0)+T(Q,-Q;2Q)$. Imposing the hard-core constraint with $U \rightarrow \infty$, we get an additional equation,
$$ 
\int\frac{T(k,k';p)}{\epsilon_{k+p}+\epsilon_{k'-p}+\Omega} \frac{dp}{2\pi}=1.
$$
Due to the form of the interaction $V(q)$, we expand the full renormalized interaction as, $T(k,k';q)=A_0 + \sum_n A_n\cos nqa$, where the coefficients $A_0, A_n$ depends on $k,k'$.  Putting this ansatz in Eq.\eqref{renor}, we get a set of coupled equations for $m>0$,
$$
A_m=\frac{2V}{m^3}-\frac{2V}{m^3}\sum^{\infty}_{m'=0}A_{m'}\int\frac{\cos mpa \cos m'pa}{\hbar^2p^2/m^{*}+\Omega}\frac{dp}{2\pi}.
$$
and from the constraint condition,
$$
A_0=\sum^{\infty}_{m=0}\int\frac{A_m\cos mpa}{\hbar^2p^2/m^{*}+\Omega}\frac{dp}{2\pi}.
$$
We found that in the limit of $\Omega \rightarrow 0$, the magnitude of the integral like $\int\frac{\cos mpa}{\hbar^2p^2/m^{*}+\Omega}\frac{dp}{2\pi}$ falls off when $m>1$. Then we get the following relation,
$$
A_m=\frac{2V}{m^3}-\frac{2V}{m^3}\sum^{m+1}_{m'=m-1}A_{m'}\int\frac{\cos (m-m')pa}{\hbar^2p^2/m^{*}+\Omega}\frac{dp}{2\pi}.
$$
We numerically find convergent solution for the $A_m$ by taking $m_{max}=100$.

\end{document}